\documentclass[aps,prd,twocolumn,sgroupedaddress]{revtex4-2}

\usepackage{amssymb}
\usepackage{amsmath}
\usepackage{graphicx}
\usepackage{enumerate}
\usepackage{mathtools}
\usepackage{color}
\usepackage{bbold}
\usepackage{subfigure}
\usepackage{slashed}
\usepackage{xcolor}
\usepackage{bm}
\usepackage{multirow}
\usepackage{pifont}

\begin{document}


\title{Self-consistency in models of neutrino scattering and fast flavor conversion}


\author{Lucas Johns}
\email[]{NASA Einstein Fellow (ljohns@berkeley.edu)}
\affiliation{Departments of Astronomy and Physics, University of California, Berkeley, CA 94720, USA}

\author{Hiroki Nagakura}
\affiliation{Division of Science, National Astronomical Observatory of Japan, 2-21-1, Osawa, Mitaka, Tokyo 181-8588, Japan}

\begin{abstract}
Several recent numerical studies have examined the effects of neutrino neutral-current scattering on fast flavor conversion (FFC). Those studies are in apparent conflict, with some finding enhancement of flavor conversion and others finding suppression. The ones that report enhancement all use homogeneous models, and we discuss in detail the self-consistency issues that they face as a result. We reproduce scattering-enhanced FFC in our own homogeneous calculations, showing that it occurs for both neutral- and charged-current scattering and that it is due to isotropization of the angular distributions over time. Because this is the very feature of the calculations that is not self-consistent, we conclude that the enhancement effect is of unclear astrophysical relevance and may not occur in natural environments.
\end{abstract}

\maketitle

\section{Introduction}

One of the challenges of simulating core-collapse supernovae is faithfully capturing the transition of neutrinos from trapping to transparency. During the first second after core bounce, the mantle of the proto-neutron star (PNS) contracts and the density profile steepens. While this process is ongoing, neutrino decoupling occurs over a region with significant radial extent. The opaque--transparent boundary becomes sharp only once the mantle has retreated to form a well-defined PNS edge. A similar challenge faces transport in neutron-star mergers, with the added complexity of emission being sourced by both the accretion disk and (when a hypermassive neutron star is formed) the central remnant.

Most of the literature on neutrino flavor transformation in these environments has focused on transparent regions, where the kinetic evolution is in the coherent (\textit{i.e.}, collisionless) regime. But it has also been recognized that in some cases the kinetics may need to be tracked in the collisional regime, where oscillations, refraction, and collisions are all dynamically important \cite{stodolsky1987, sigl1993, mckellar1994, strack2005, vlasenko2014, volpe2015, kartavtsev2015, blaschke2016, richers2019}. While the phenomenology of this regime is precisely understood in the early universe \cite{savage1991, lunardini2001, dolgov2002, abazajian2002, wong2002, mangano2002, mangano2005, pastor2009, gava2010, mangano2011, mangano2012, castorina2012, grohs2016, grohs2017, johns2016, johns2018, barenboim2017, froustey2021}, it is much less so in compact objects \cite{cherry2012, sarikas2012b, cirigliano2018, cherry2020}.

Interest in the collisional regime has been invigorated by the study of fast instabilities, which are associated with angular crossings in the neutrino distributions \cite{sawyer2005, sawyer2009, sawyer2016, chakraborty2016b, chakraborty2016c, tamborra2017, dasgupta2017, wu2017, wu2017b, dasgupta2018, dasgupta2018b, airen2018, abbar2018, abbar2019, capozzi2019, capozzi2019b, yi2019, abbar2019b, azari2019, shalgar2019, nagakura2019, chakraborty2020, martin2020, johns2020, johns2020b, bhattacharyya2020, abbar2020, shalgar2020b, capozzi2020, xiong2020, shalgar2020, bhattacharyya2020b, morinaga2020, glas2020, abbar2020b, padilla2020, george2020, johns2021, nagakura2021, nagakura2021b, nagakura2021c, padillagay2022, roggero2022, abbar2022, harada2022}. Because angular distributions are shaped by collisional processes, fast instabilities and collisions are naturally related, and simulation data indicate that fast instabilities probably occur in locations where collisions are not entirely negligible. Thus Refs.~\cite{capozzi2019, shalgar2019} considered how collisions set up angular crossings and modify the development of fast flavor conversion (FFC), and Refs.~\cite{johns2021b, dasgupta2021, martin2021} conducted linear stability analyses with nonvanishing collisional terms.

Recent works have especially focused on neutral-current (NC) scattering and its effects on FFC. Shalgar and Tamborra \cite{shalgar2021} reported the enhancement of flavor conversion, rather than its damping, in homogeneous calculations. These numerical findings were replicated by Kato, Nagakura, and Morinaga \cite{kato2021}, analyzed by Sasaki and Takiwaki \cite{sasaki2021}, and extended by Hansen, Shalgar, and Tamborra \cite{hansen2022}. In contrast, Martin, Carlson, Cirigliano, and Duan \cite{martin2021}, running inhomogeneous computations, found only damping. Sigl \cite{sigl2022}, also solving an inhomogeneous model, likewise observed no enhancement effect.

Our goal is to achieve a better understanding of the apparently conflicting results of Refs.~\cite{shalgar2021, martin2021, sigl2022, kato2021, sasaki2021, hansen2022}. We do not attempt to develop a quantitative theory that accounts for the enhancement effect of Refs.~\cite{shalgar2021, kato2021, sasaki2021, hansen2022}, which we henceforth refer to as \textit{scattering-enhanced FFC}. Doing so would likely be challenging, as the nonlinear theory of FFC \cite{johns2020, johns2020b, bhattacharyya2020, bhattacharyya2021, padillagay2021, duan2021} is difficult enough on its own without the complication of scattering. We opt instead for a simpler approach, comparing numerical results obtained using different collisional terms.

We favor this approach because the models exhibiting scattering-enhanced FFC face self-consistency issues. Being homogeneous, they inevitably lead to isotropic neutrino distributions at late times. The inconsistency here is between this late-time isotropy and the initial-time anisotropy. The transition from the latter to the former occurs even if the oscillation Hamiltonian $H$ is turned off. With such significant evolution occurring independently of oscillations, it is not immediately clear how to interpret the evolution that occurs when $H$ is turned on. Self-consistency, by our definition, means that the initial state is constant under the $H=0$ equations of motion. When this fails to be the case, the system exhibits transient behavior having nothing to do with oscillations.

The results of our comparison test are illuminating. First we replicate scattering-enhanced FFC. Comparing NC and charged-current (CC) scattering, we find almost identical enhancements. This rules out any special connection between the enhancement effect and the particular character of NC scattering. We then compare to CC scattering \textit{without} isotropization of the angular distributions and find that the enhancement effect disappears. The crucial role of isotropization is confirmed by closely inspecting the evolution in the isotropizing calculations: scattering-enhanced FFC is clearly a cumulative effect of gradually changing the angular distributions over time \cite{hansen2022}. This latter feature, however, is unrealistic. It is a product of using a model that is not fully self-consistent, as mentioned above and further explained in the next section. The implication is that scattering-enhanced FFC, at least in the form observed in Refs.~\cite{shalgar2021, kato2021, sasaki2021, hansen2022}, is liable to disappear upon moving to a more realistic model.

Incidentally, another work by one of us makes the claim that collisional instabilities associated with unequal neutrino and antineutrino interaction rates are capable of not only enhancing flavor conversion but causing it to occur where essentially none would otherwise \cite{johns2021b}. Collisional instabilities are distinct phenomena from scattering-enhanced FFC. They are an instability class in their own right and can occur in the absence of slow or fast instabilities. Moreover, they are not undermined by self-consistency issues like scattering-enhanced FFC is. In this study, though, the focus is on models with \textit{equal} neutrino and antineutrino interaction rates, making collisional instabilities irrelevant to the calculations presented.

In Sec.~\ref{sec:interpretive} we elaborate on some of the subtleties involved in examining collisional neutrino flavor transformation. In Sec.~\ref{sec:eoms} we specify the model we use for the numerical comparison test, which is meant to disentangle what is important for scattering-enhanced FFC and what is not. In Sec.~\ref{sec:results} we present the results. In Sec.~\ref{sec:discussion} we close with a few brief remarks.

\section{Modeling \label{sec:interpretive}}

In this section we describe the self-consistency issues faced by homogeneous models (Sec.~\ref{sec:selfcons}), the different scattering implementations we will be comparing (Sec.~\ref{sec:scattpheno}), and the important distinction between having an isotropic gain term and having actual isotropization as a function of time (Sec.~\ref{sec:isoiso}).

\subsection{Self-consistency \label{sec:selfcons}}

It is impossible to self-consistently treat anisotropy and collisions in a homogeneous model of the neutrino flavor field. In compact-object environments, the anisotropy of the angular distributions is the result of a negotiation between advection and inhomogeneity (specifically the global geometry of the environment) on the one hand and collisions on the other. The advection of neutrinos away from the center of a supernova, for example, causes angular distributions to become more forward-peaked. Collisions cause distributions to become more isotropic. In regions of intermediate opacity, distributions are shaped by both factors. Loosely speaking,
\begin{equation}
\textrm{collisions} + \textrm{inhomogeneity} = \textrm{anisotropy}. \notag
\end{equation}
Interpretive issues arise when we try to draw conclusions from calculations combining anisotropy, collisions, and \textit{homogeneity}, because one term in the equation is then missing.

What we mean by self-consistency here is that the initial state should be a steady state of the classical (non-oscillating) equations of motion. A homogeneous and \textit{collisionless} calculation is self-consistent according to this definition because the classical solution shows no evolution. Any flavor development that occurs when oscillations are turned on is therefore a consequence specifically of oscillations. But in a homogeneous and \textit{collisional} calculation, collisions inevitably isotropize the neutrino angular distributions. Part of the dynamics, then, is simply due to starting away from the classical steady state.

Ref.~\cite{shalgar2021}, which first reported enhancement of FFC due to scattering, admitted the lack of self-consistency in their treatment of the problem. A calculation does not need to be fully self-consistent to be informative. The inconsistency, though, will hopefully not be pivotal to the results. If it is, then the results may be \textit{attributable} to the inconsistency, which would call into question the physical relevance of the findings. In this paper we ask whether scattering-enhanced FFC is undermined in this manner.

\subsection{Scattering phenomenology \label{sec:scattpheno}}

Our chosen tactic is to vary the implementation of scattering, assuming homogeneity throughout. In doing so, we never evade the self-consistency problem, but we are able to determine which aspects of the scattering physics are relevant to the enhancement of FFC and which are not.

Collisions generally damp and suppress oscillations. The measurement paradigm accounts for this tendency in an intuitive way. In each collision, a background particle measures or observes the test neutrino. A measurement of flavor decoheres the quantum state in the flavor basis, much as a measurement of position localizes a wave function in space. At the extreme, rapid collisions cause the quantum Zeno effect, wherein the wave function is stuck in its initial state, unable to evolve. This effect has been documented and examined many times in the context of particle mixing \cite{stodolsky1987, mckellar1994, abazajian2001b, kishimoto2008, johns2019, johns2019b, kishimoto2020, johns2020c, johns2020d}.

Collisional instabilities flout the principle that collisions suppress flavor conversion \cite{johns2021b}. They are an explicable exception, though. Collisions do indeed act to decohere neutrino and antineutrino flavor states, but because they are decohered \textit{differently}---that is, at unequal rates---there is a differential effect that feeds into the self-coupling term in the Hamiltonian. This effect destabilizes the flavor field, enhancing flavor conversion beyond what would be expected of a stable system.

Scattering-enhanced FFC presumes to be another exception. To date, all of the studies exhibiting the phenomenon employ NC scattering, which does not adhere to the measurement paradigm. Neutrino--nucleon scattering, for example, transfers momentum to or from a neutrino without measuring the flavor state. This process is coherent at the single-particle level in the sense that flavor information is preserved as the particle is moved from one momentum to another. From these observations it follows that scattering-enhanced FFC cannot be rooted in flavor decoherence due to system--environment interaction, as collisional instability is, but rather must be rooted in the redistribution of neutrino momentum.

It is natural to go further and wonder whether there is something special about NC scattering in particular---as opposed to charged-current (CC) scattering, for example---that causes it to interact with oscillations in unexpected ways. This is not necessarily entailed by past studies of scattering-enhanced FFC, since none of them compare NC and CC scattering, and it could very well be that the mechanism responsible for enhancing flavor conversion is common to both types. Nonetheless, it is an intriguing possibility.

Moreover, there is reason to expect the phenomenology associated with NC scattering to differ from that of CC scattering. To see this mathematically, we now introduce notation pertaining to the implementation of collisions. As in Ref.~\cite{johns2021b}, we apply a relaxation-time approximation to each process individually. With the different processes indexed by $P$, the full collision term is
\begin{equation}
i \mathcal{C} = \sum_P \lbrace \Gamma_P, \rho_P - \rho \rbrace,
\end{equation}
where $\Gamma_P = \textrm{diag} ( \Gamma_e^P / 2, \Gamma_x^P / 2)$ is a diagonal matrix whose entries are the $\nu_e$ and $\nu_x$ relaxation rates due to process $P$. It is not only the interaction rates that are process-dependent. The flavor states $\rho_P$ toward which the system collisionally relaxes are as well. 

The prescription for NC scattering is the same as used in Refs.~\cite{shalgar2021, kato2021, sasaki2021, hansen2022}:
\begin{equation}
\textbf{\textrm{NC (isotropizing):}} ~~~ i\mathcal{C} =  \Gamma \left( \rho_\textrm{NC} - \rho \right). \label{eq:nciso}
\end{equation}
where
\begin{equation}
\rho_\textrm{NC} = \langle \rho \rangle = 
\begin{pmatrix}
\langle \rho \rangle_{ee} & \langle \rho \rangle_{ex} \\
\langle \rho \rangle_{xe} & \langle \rho \rangle_{xx}
\end{pmatrix} \label{eq:ncisoeq}
\end{equation}
and $\langle \dots \rangle$ represents an average over momentum. This mocks up neutrino--nucleon scattering: flavor is not measured during the collision, it is merely redirected isotropically and redistributed over energy. Because NC scattering is flavor-blind, there is only one relaxation rate.

Crucially, $\rho_\textrm{NC}$ is a \textit{coherent} average in the same sense that the Hamiltonian from neutrino--neutrino forward scattering is. As a result, collective effects stemming from NC scattering are possible, much as collective effects stemming from the Hamiltonian are \cite{bell2001}. This point motivates further examination of the possibility that scattering-enhanced FFC is uniquely related to NC scattering.

Another possibility is that scattering-enhanced FFC is common to both NC and CC scattering. If so, then it must be a result of shuffling number densities (the diagonal entries of the density matrices) across momentum bins, not a result of shuffling flavor states (the diagonal \textit{and} off-diagonal entries). 

Based on past studies, we are unsure which of these possibilities is correct, \textit{i.e.}, whether scattering-enhanced FFC is uniquely tied to the coherent character of NC scattering or is a product of isotropizing the number densities. The easy way to rule out the first possibility is to compare a calculation with NC scattering to one with CC scattering. In the latter case, the prescription for the collision term is
\begin{equation}
\textbf{\textrm{CC (isotropizing):}} ~~~ i\mathcal{C} =  \Gamma \left( \rho_\textrm{CC}^\textrm{iso} - \rho \right) \label{eq:cciso}
\end{equation}
where
\begin{equation}
\rho_\textrm{CC}^\textrm{iso} = 
\begin{pmatrix}
\langle \rho \rangle_{ee} & 0 \\
0 & \langle \rho \rangle_{xx}
\end{pmatrix}. \label{eq:ccisoeq}
\end{equation}
This mocks up, for example, CC neutrino--electron scattering: flavor is measured during the collision \textit{and} it is redirected isotropically. Decoherence means that CC scattering does not source a coherent average.

\begin{table}
\centering
\begin{tabular}{|c||c|c|}
 \hline
 \multicolumn{3}{|c|}{Effects on a single scattered neutrino} \\
 \hline\hline
 ~~Implementation~~ & ~~Changes $\mathbf{p}$~~ & ~~Changes flavor~~ \\
 \hline
None & \ding{55} & \ding{55} \\
NC (iso) & \ding{51} & \ding{55} \\
CC (iso) & \ding{51} & \ding{51}  \\
CC (non) & \ding{55} & \ding{51} \\
 \hline
\end{tabular}
\caption{Comparison of the effects of the different scattering implementations used in this paper, considered at the single-particle level. The more rigorous meanings of the columns are found in the relevant equations in Sec.~\ref{sec:scattpheno}. Note that non-isotropizing NC scattering has no effect in the single-energy case. \label{tablescatter}}
\end{table}

Table~\ref{tablescatter} identifies the differences between the \textit{NC (isotropizing)} and \textit{CC (isotropizing)} prescriptions.

We fill out the table with two other implementations of scattering. One of them is
\begin{equation}
\textbf{\textrm{CC (non-isotropizing):}} ~~~ i\mathcal{C} =  \Gamma \left( \rho_\textrm{CC}^\textrm{non} - \rho \right) \label{eq:ccnon}
\end{equation}
where
\begin{equation}
\rho_\textrm{CC}^\textrm{non} = 
\begin{pmatrix}
\rho_{ee} & 0 \\
0 & \rho_{xx}
\end{pmatrix}. \label{eq:ccnoneq}
\end{equation}
This collisional term can also be written as
\begin{equation}
i\mathcal{C} =  - \Gamma \rho_T,
\end{equation}
where
\begin{equation}
\rho_T = \begin{pmatrix}
0 & \rho_{ex} \\
\rho_{xe} & 0
\end{pmatrix},
\end{equation}
which makes it clear that non-isotropizing CC scattering is the usual damping prescription. It describes a process that measures flavor but does not alter momentum. It is not important that this collisional form be physically motivated. By isolating decoherence, it will help us to confirm that scattering-enhanced FFC is related to redistributing momentum and not to damping coherence.

The final entry corresponds to the collisionless case, which of course has
\begin{equation}
\textbf{\textrm{None:}} ~~~ i\mathcal{C} =  0.
\end{equation}
For a single-energy calculation, \textit{None} can also be regarded as signifying non-isotropizing NC scattering, which in this case does nothing. Thinking about it this way adds a certain degree of symmetry to Table~\ref{tablescatter}. More generally, non-isotropizing NC scattering would change $E_{\nu}$ without changing $\mathbf{\hat{p}}$. This type of process is not useful for the present study.

\subsection{Static vs. dynamic isotropization \label{sec:isoiso}}

In Sec.~\ref{sec:results} we will find that \textit{NC (isotropizing)} and \textit{CC (isotropizing)} both exhibit scattering-enhanced FFC but \textit{CC (non-isotropizing)} does not. The enhancement effect is evidently a result of isotropizing number densities.

This leaves open one further question: What exactly is it about isotropic scattering that is responsible for the enhancement? Here we need to distinguish between two possibilities. The first, which we call \textit{static isotropization}, refers to a situation in which the scattering gain term is isotropic but the angular distributions remain anisotropic due to inhomogeneity and neutrino advection. The second, which we call \textit{dynamic isotropization}, refers to a situation in which an isotropic gain term causes angular distributions to isotropize as a function of time. 

In an inhomogeneous environment, scattering can be isotropic---that is, sourcing neutrinos identically in all directions---without actually isotropizing the angular distributions over time. Homogeneity collapses these two distinct aspects of the physics into one. Without the countervailing influence of neutrino advection, isotropic scattering implies isotropization. It is important that we disentangle the two, because while (quasi-)isotropic scattering is realistic, local isotropization in time is not. Angular distributions at a given location in a supernova or merger do evolve, but this evolution is driven by global changes in the environment rather than local collisional processes. 

We will see in Sec.~\ref{sec:results} that the evolution in our homogeneous calculations hints at dynamic isotropization being the key to scattering-enhanced FFC, as the flavor conversion tracks the changing angular distributions. This result suggests in turn that scattering-enhanced FFC is a transient feature of homogeneous calculations. It is our assessment that there is no evidence at present for scattering-enhanced FFC occurring in real core-collapse supernovae or neutron-star mergers.

\section{Equations of motion \label{sec:eoms}}

We assume spatial homogeneity, axial symmetry, and the single-energy approximation. The neutrino and antineutrino flavor density matrices $\rho_v$ and $\bar{\rho}_v$, for propagation angle $v = \cos \theta$, thus obey the equations of motion
\begin{align}
&i \frac{d}{dt} \rho_v =  \left[ + H^\textrm{vac} + H^{\nu\nu}_v, \rho_v \right] + i C_v [\rho] \notag \\
&i \frac{d}{dt} \bar{\rho}_v = \left[ - H^\textrm{vac} + H^{\nu\nu}_v, \bar{\rho}_v \right] + i C_v [\bar{\rho}].
\end{align}
The vacuum Hamiltonian is
\begin{equation}
H_\textrm{vac} = \frac{\omega}{2}
\begin{pmatrix}
- \cos 2\theta & \sin 2\theta \\
\sin 2\theta & \cos 2\theta
\end{pmatrix} 
\end{equation}
where $\omega = \delta m^2 / 2E_\nu$ is the vacuum oscillation frequency, $\delta m^2$ is the mass-squared splitting, $E_\nu$ is neutrino energy, and $\theta$ is the mixing angle. We use a matter-suppressed mixing angle rather than explicitly including the Hamiltonian term from neutrino--electron forward scattering. The neutrino--neutrino forward-scattering Hamiltonian is
\begin{equation}
H_v^{\nu\nu} = \sqrt{2} G_F \int_{-1}^{+1} d v' ( 1 - v v' ) ( \rho_{v'} - \bar{\rho}_{v'} ),
\end{equation}
where the integral is over all background neutrino momenta.

We compare three different implementations of scattering. We use Eqs.~\eqref{eq:nciso} and \eqref{eq:ncisoeq} for isotropizing NC scattering, Eqs.~\eqref{eq:cciso} and \eqref{eq:ccisoeq} for isotropizing CC scattering, and Eqs.~\eqref{eq:ccnon} and \eqref{eq:ccnoneq} for non-isotropizing CC scattering. The momentum average $\langle \dots \rangle$ is now defined as
\begin{equation}
\langle \rho \rangle = \frac{1}{2} \int_{-1}^{+1} dv\rho_v
\end{equation}
because we assume that neutrinos have uniform probability of being scattered into any propagation direction. Written out for clarity, the collision terms are thus
\begin{equation}
i\mathcal{C}_v = \Gamma \begin{pmatrix}
\langle \rho \rangle_{ee} - \rho_{v,ee} & \langle \rho \rangle_{ex} - \rho_{v,ex} \\
\langle \rho \rangle_{xe} - \rho_{v,xe}& \langle \rho \rangle_{xx} - \rho_{v,xx}
\end{pmatrix} \label{eq:ncengy}
\end{equation}
for isotropizing NC scattering,
\begin{equation}
i\mathcal{C}_v = \Gamma \begin{pmatrix}
\langle \rho \rangle_{ee} - \rho_{v,ee} & - \rho_{v,ex} \\
- \rho_{v,xe}& \langle \rho \rangle_{xx} - \rho_{v,xx}
\end{pmatrix} \label{eq:ccengy}
\end{equation}
for isotropizing CC scattering, and
\begin{equation}
i\mathcal{C}_v = \Gamma \begin{pmatrix}
0 & - \rho_{v,ex} \\
- \rho_{v,xe} & 0
\end{pmatrix} \label{eq:ccengy}
\end{equation}
for non-isotropizing CC scattering. Expressions for antineutrinos are obtained by replacing $\rho \rightarrow \bar{\rho}$.

For the calculations, we adopt a set of parameters very similar to those used in Ref.~\cite{johns2021b}. The idea is to pick values representative of the neutrino decoupling region during the supernova accretion phase. The parameters are given in Table~\ref{tablevparams}. We note again that a matter-suppressed mixing angle is used in place of the vacuum mixing angle. Its value is chosen to be consistent with the matter potential implied by the electron density. The normal mass ordering is assumed.

\begin{table*}
\centering
\begin{tabular}{|c|c|c|}
 \hline
 \multicolumn{3}{|c|}{Parameters used in Sec.~\ref{sec:results} calculations} \\
 \hline\hline
 ~~Variable~~ & ~~Meaning~~ & ~~Value~~ \\
 \hline
$\rho$ & Fluid density & $10^{12}$~g/cm$^3$ \\
$T$ & Fluid temperature & 7 MeV \\
$\mu_e$ & Electron chemical potential & 20 MeV \\
$Y_e$ & Electron fraction & 0.13 \\
$E_\nu$ & Neutrino energy & 20 MeV \\
~~~~$n_{\nu_e,0} (0)$~~~~ & ~~~~$l=0$ Legendre moment of $n_{\nu_e,v} (0)$~~~~ & ~~~~$3.0 \times 10^{33}$~cm$^{-3}$~~~~ \\
$n_{\nu_e,1} (0)$ & $l=1$ Legendre moment of $n_{\nu_e,v} (0)$ & $1.5 \times 10^{32}$~cm$^{-3}$ \\
$n_{\nu_e,2} (0)$ & $l=2$ Legendre moment of $n_{\nu_e,v} (0)$ & $6.0 \times 10^{30}$~cm$^{-3}$ \\
$n_{\bar{\nu}_e,0} (0)$ & $l=0$ Legendre moment of $n_{\bar{\nu}_e,v} (0)$ & $2.5 \times 10^{33}$~cm$^{-3}$ \\
$n_{\bar{\nu}_e,1} (0)$ & $l=1$ Legendre moment of $n_{\bar{\nu}_e,v} (0)$ & $2.5 \times 10^{32}$~cm$^{-3}$ \\
$n_{\bar{\nu}_e,2} (0)$ & $l=2$ Legendre moment of $n_{\bar{\nu}_e,v} (0)$ & $2.0 \times 10^{31}$~cm$^{-3}$ \\
$n_{\nu_x,0} (0)$ & $l=0$ Legendre moment of $n_{\nu_x,v} (0)$ & $1.0 \times 10^{33}$~cm$^{-3}$ \\
$n_{\nu_x,1} (0)$ & $l=1$ Legendre moment of $n_{\nu_x,v} (0)$ & $1.5 \times 10^{32}$~cm$^{-3}$ \\
$n_{\nu_x,2} (0)$ & $l=2$ Legendre moment of $n_{\nu_x,v} (0)$ & $1.7 \times 10^{31}$~cm$^{-3}$ \\
$\delta m^2$ & Mass-squared splitting & $2.4 \times 10^{-3}$~eV$^2$ \\
$\theta$ & Mixing angle & $10^{-6}$ \\
$\omega$ & Vacuum oscillation frequency & 0.3 km$^{-1}$ \\
$\mu | \mathbf{D}_0 (0) |$ & Neutrino--neutrino forward-scattering potential & $3 \times 10^5$ km$^{-1}$ \\ 
$\Gamma$ & Scattering rate & 5.2 km$^{-1}$ \\ 
 \hline
\end{tabular}
\caption{Fixed parameters used in the calculations. $n_{\nu_\alpha,v}(0)$ is the initial number density of $\nu_\alpha$ with propagation angle $v$ and can be constructed from the Legendre moments given in the table. The neutrino--neutrino forward scattering potential $\mu | \mathbf{D}_0 (0)|$ is written in terms of the usual self-coupling parameter $\mu$ and the initial-time $l=0$ difference vector $\mathbf{D}_0$. It is equivalent to $\sqrt{2} G_F [ n_{\nu_e} (0) - n_{\bar{\nu}_e} (0) ]$. \label{tablevparams}}
\end{table*}

\begin{figure}
\includegraphics[width=.42\textwidth]{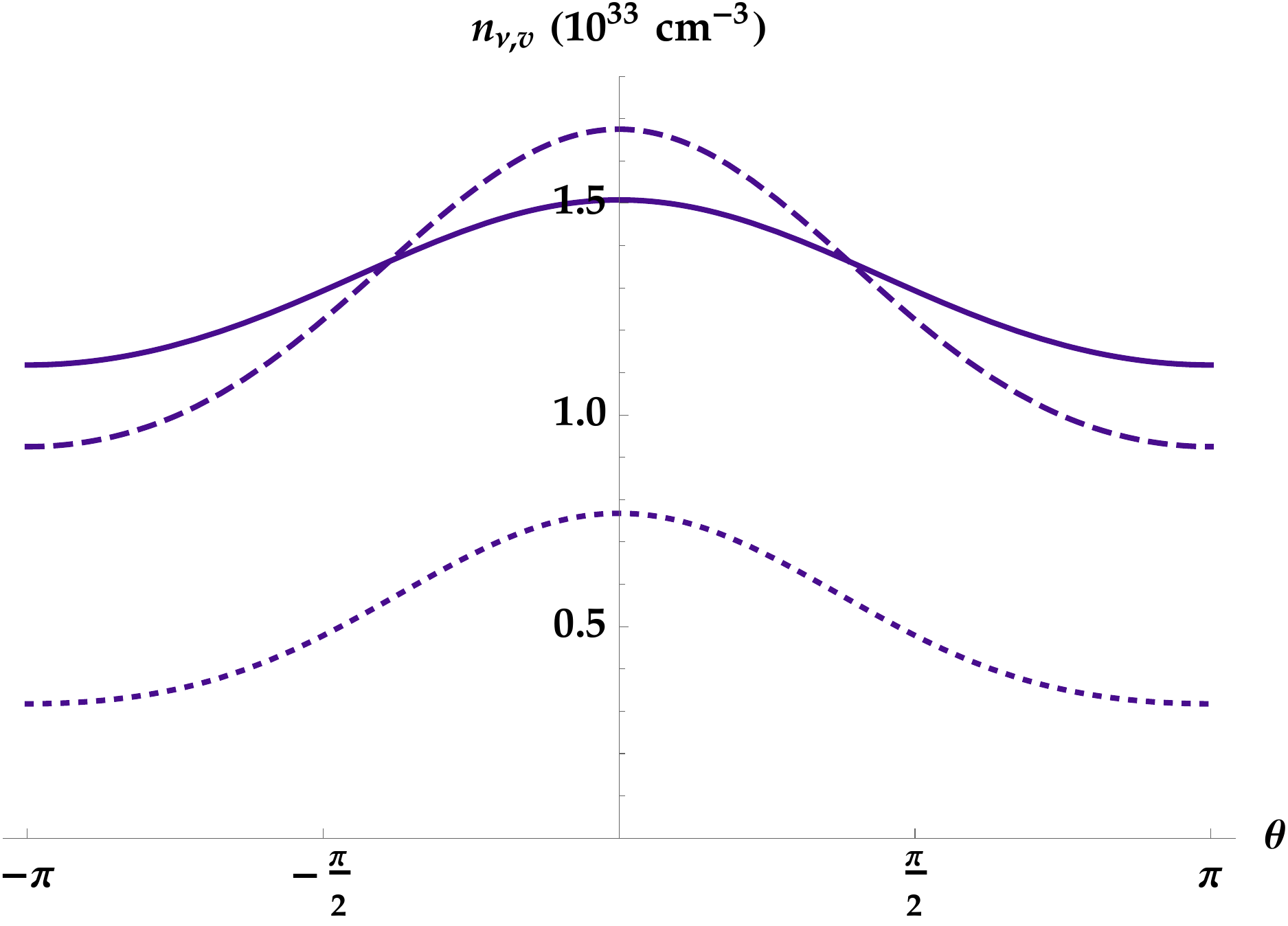}
\caption{Angular distributions used in the Sec.~\ref{sec:results} calculations: $n_{\nu_e, v}$ (solid), $n_{\bar{\nu}_e, v}$ (dashed), $n_{\nu_x, v} = n_{\bar{\nu}_x, v}$ (dotted), where $v = \cos\theta$ and $\theta$ is the neutrino propagation angle. \label{ang_distrs}}
\end{figure}

For the angular distributions we use
\begin{equation}
n_{\nu_\alpha, v} = \frac{1}{2} n_{\nu_\alpha, 0} L_0 (v) + \frac{3}{2} n_{\nu_\alpha, 1} L_1 (v) + \frac{5}{2} n_{\nu_\alpha, 2} L_2 (v),
\end{equation}
where $n_{\nu_\alpha, l}$ is the $l$th Legendre moment of the angular distribution of flavor $\alpha$ and $L_l (v)$ is the $l$th Legendre polynomial. By definition, $n_{\nu_\alpha, 0} = n_{\nu_\alpha}$. We pick
\begin{gather}
n_{\nu_e, 1} = 0.050 n_{\nu_e}, ~~~ n_{\nu_e, 2} = 0.002 n_{\nu_e}, \notag \\
n_{\bar{\nu}_e, 1} = 0.100 n_{\bar{\nu}_e}, ~~~ n_{\bar{\nu}_e, 2} = 0.008 n_{\bar{\nu}_e}, \notag \\
n_{\nu_x, 1} = 0.150 n_{\nu_x}, ~~~ n_{\nu_x, 2} = 0.017 n_{\nu_x}.
\end{gather}
These are relatively isotropic distributions, and all moments $l > 2$ are assumed to vanish. The distributions are plotted in Fig.~\ref{ang_distrs}, and the explicit values of the moments are shown in Table~\ref{tablevparams}.

Using the values above, the rate of neutrino--neutron scattering is approximately \cite{burrows2006}
\begin{equation}
\Gamma_{\nu n} \sim \frac{n_n \sigma_0}{4} \left( \frac{1 + 3 g_A^2}{4} \right) \left( \frac{E_\nu}{m_e} \right)^2  \sim 0.52 ~\textrm{km}^{-1},
\end{equation}
as used in Ref.~\cite{johns2021b}. Here $n_n$ is the neutron density, $m_e$ is the electron rest mass, $g_A \cong -1.28$ is the axial-vector coupling constant, and $\sigma_0 = 4 G_F^2 m_e^2 / \pi$. This estimate ignores corrections from inelasticity, recoil, and weak magnetism. As described in Ref.~\cite{johns2021b}, other NC processes occur, including NC scattering on protons, four-neutrino processes, $\nu\bar{\nu}$ annihilation to $e^+ e^-$, nucleon--nucleon bremsstrahlung, and the flavor-blind part of neutrino--electron scattering. These, however, are all subdominant to neutrino--nucleon scattering.

We take a collisional rate 10 times as large,
\begin{equation}
\Gamma = 10 \Gamma_{\nu n},
\end{equation}
hence the value in Table~\ref{tablevparams}. The exaggerated scattering rate allows for convergence to be achieved at lower computational cost, and creates a larger separation between the time scales $\Gamma^{-1}$ and $\omega^{-1}$. It does not change any of the qualitative conclusions. In general, though, there is some dependence of the flavor conversion on the particular value of $\Gamma$, and in the limit of very large $\Gamma$, flavor conversion vanishes because scattering isotropizes the angular distributions and wipes out the angular crossing before fast oscillations even begin.

Clearly the use of a single relaxation rate $\Gamma$ is unrealistic for CC scattering, since $\nu_e$ and $\nu_x$ have different interaction rates in supernovae and mergers. But our present goal is to understand which parts of the NC implementation are important for the results observed in prior studies. Using a single, fixed rate for all interactions is the sensible choice for this purpose.

\section{Results \label{sec:results}}

By comparing \textit{NC (isotropizing)} to \textit{CC (isotropizing)}, we test whether the flavor-blind character of neutral currents is important. We find that it is not, as the results in these cases are virtually identical.

Ideally, we would like to compare \textit{NC (isotropizing)} to a non-isotropizing implementation of NC scattering, so as to learn more about the role of isotropization, but the latter would have no physical effect: it would neither measure flavor nor redirect momentum. What we can do instead is compare \textit{CC (isotropizing)} to \textit{CC (non-isotropizing)}. From this comparison we find that isotropization is essential for the claimed enhancement effect.

As discussed in Sec.~\ref{sec:isoiso}, this is where we reach the limits of what can be done with homogeneous modeling. It is not possible, using a homogeneous model, to discern between isotropic scattering, which in an astrophysical setting is locally balanced by advection, and isotropization over time.

\begin{figure*}
\centering
\begin{subfigure}{
\centering
\includegraphics[width=.310\textwidth]{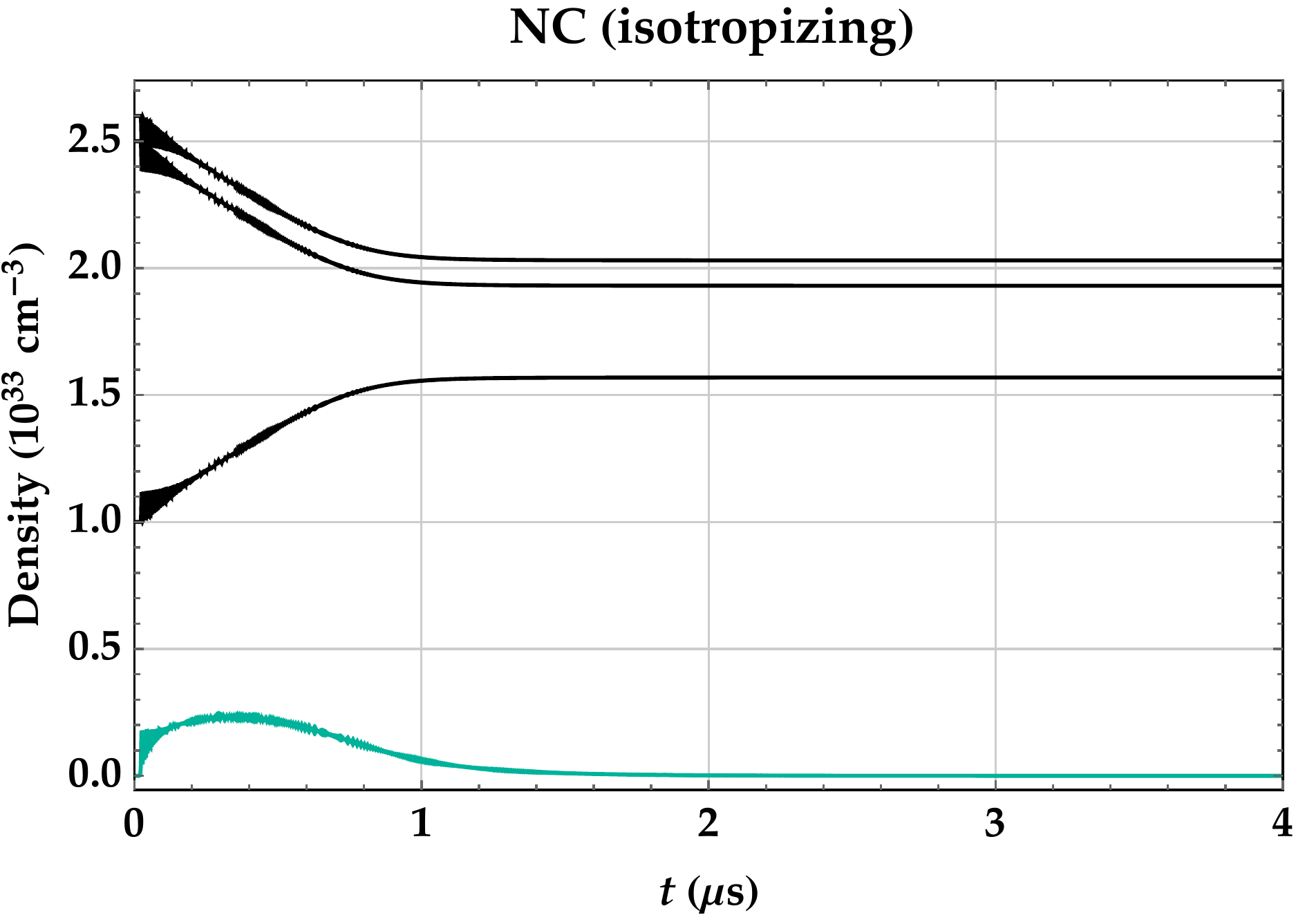}
}
\end{subfigure}
\begin{subfigure}{
\centering
\includegraphics[width=.310\textwidth]{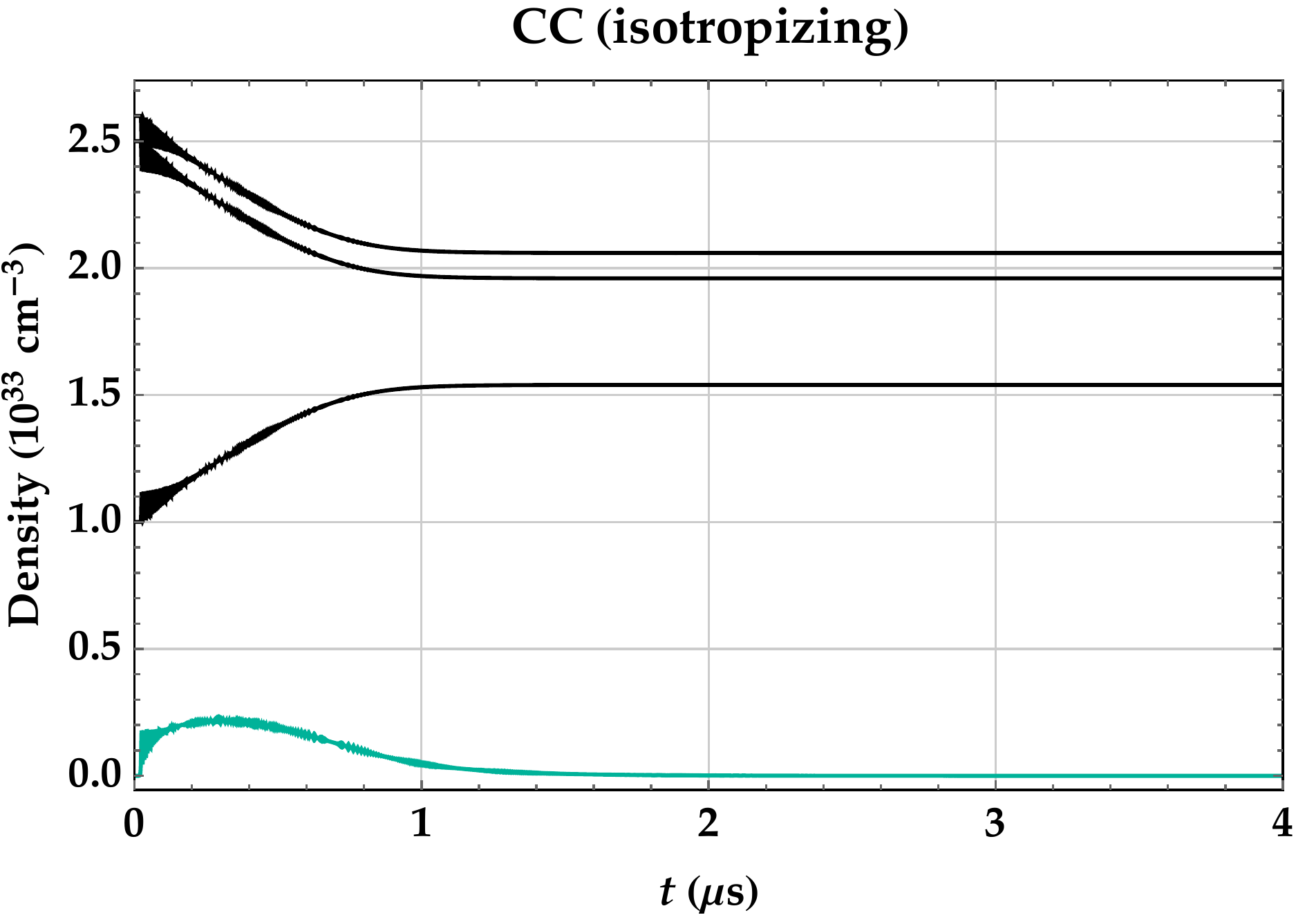}
}
\end{subfigure}
\begin{subfigure}{
\centering
\includegraphics[width=.310\textwidth]{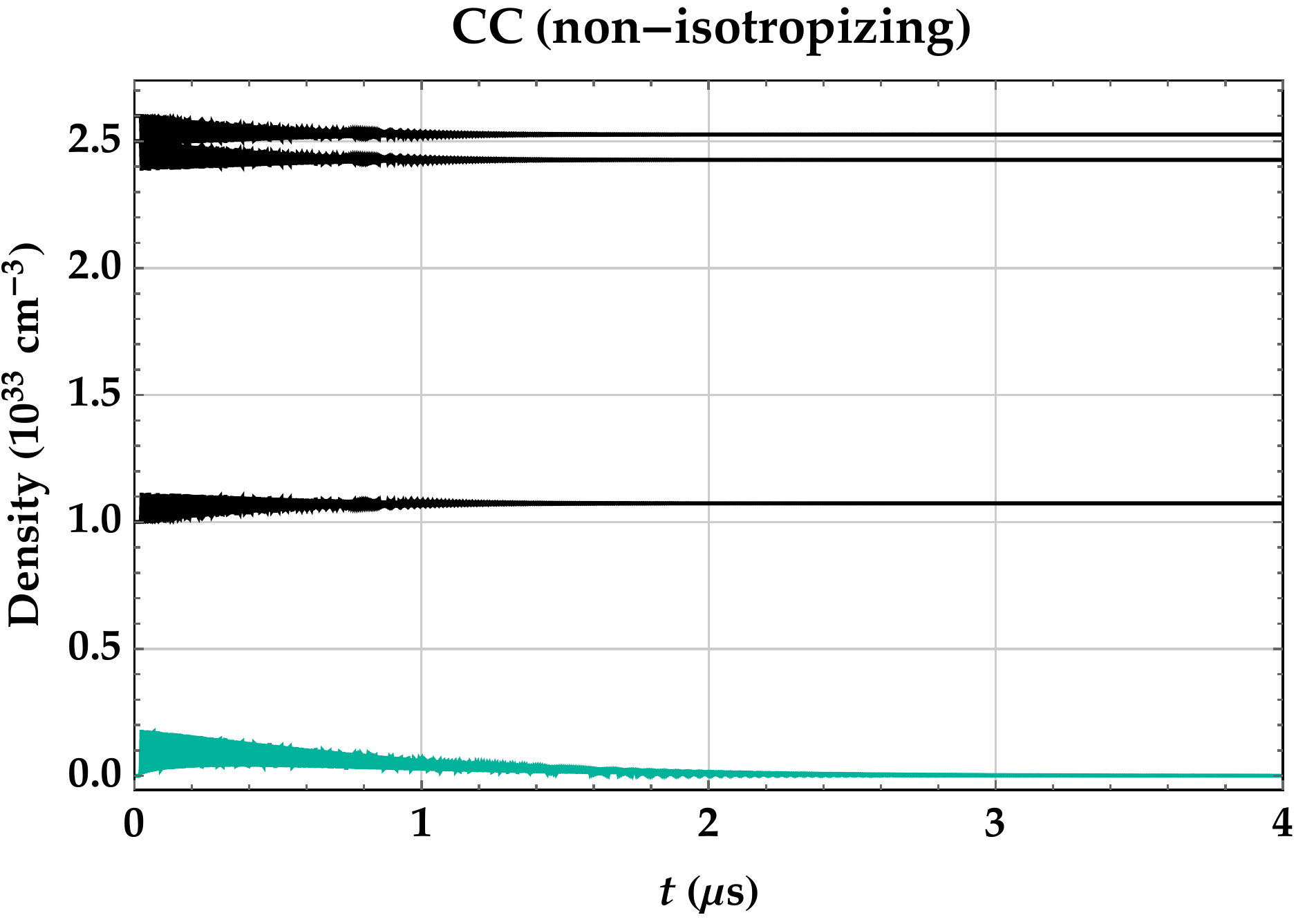}
}
\end{subfigure}

\begin{subfigure}{
\centering
\includegraphics[width=.310\textwidth]{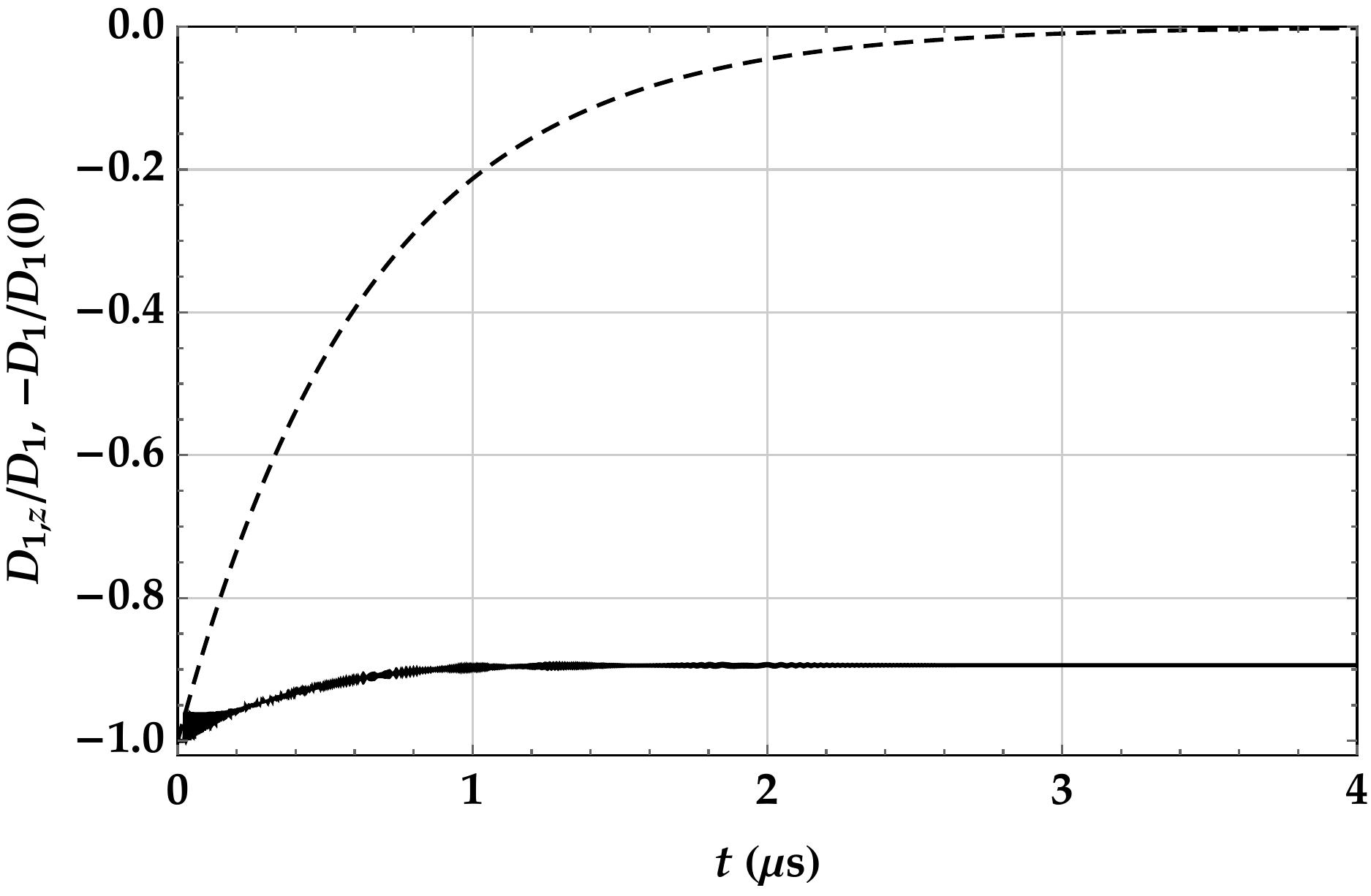}
}
\end{subfigure}
\begin{subfigure}{
\centering
\includegraphics[width=.310\textwidth]{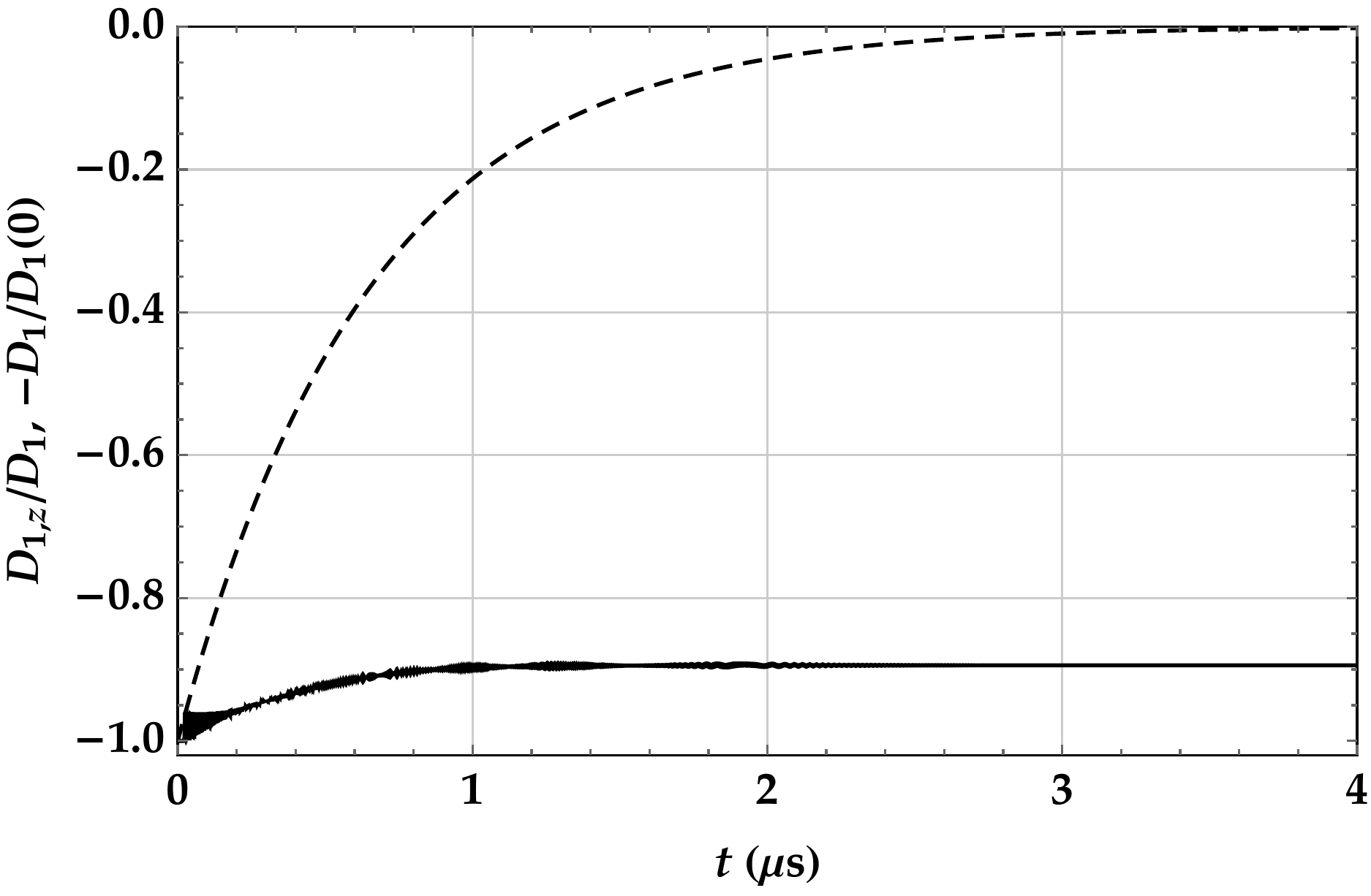}
}
\end{subfigure}
\begin{subfigure}{
\centering
\includegraphics[width=.310\textwidth]{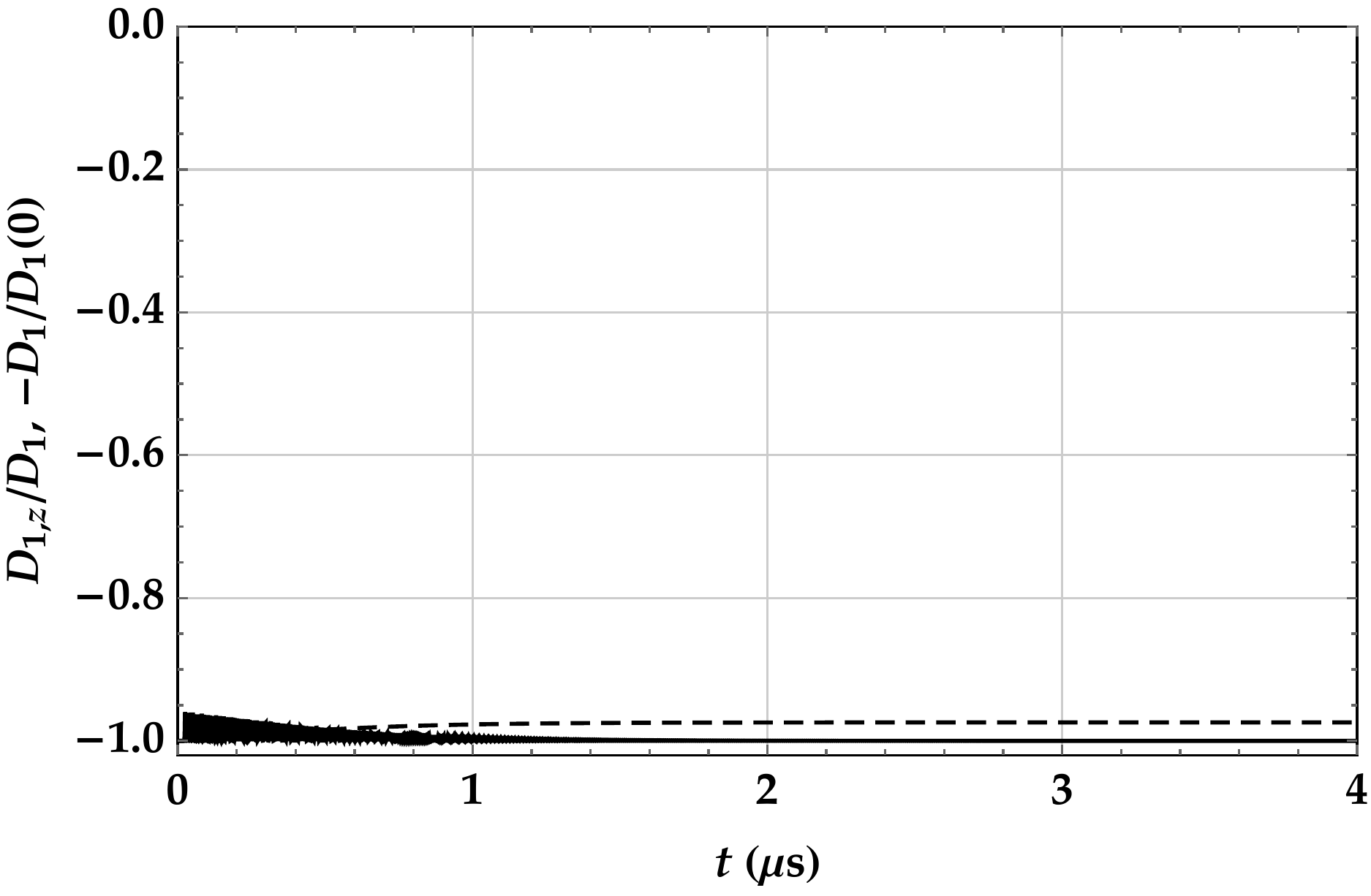}
}
\end{subfigure}

\begin{subfigure}{
\centering
\includegraphics[width=.310\textwidth]{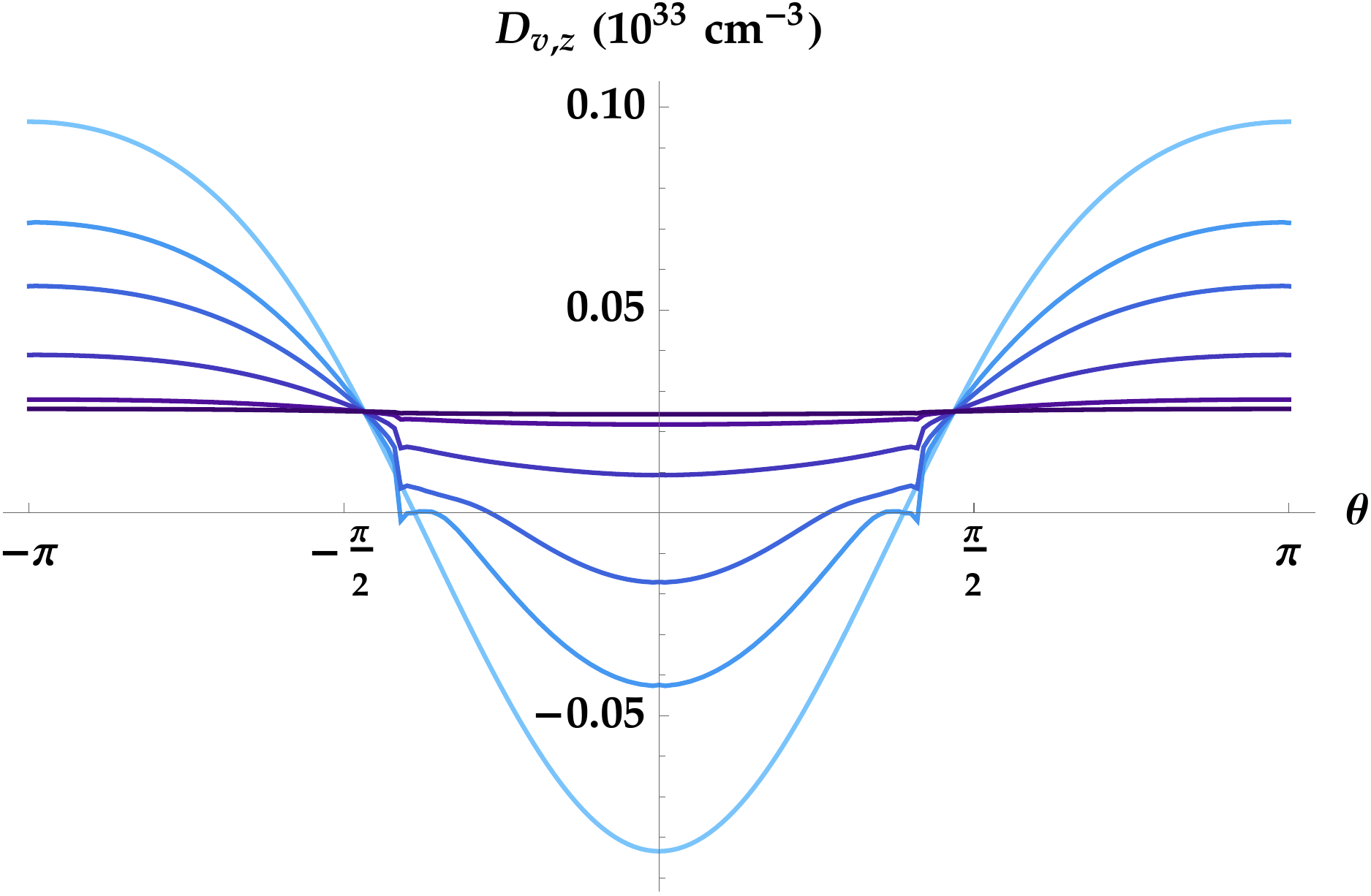}
}
\end{subfigure}
\begin{subfigure}{
\centering
\includegraphics[width=.310\textwidth]{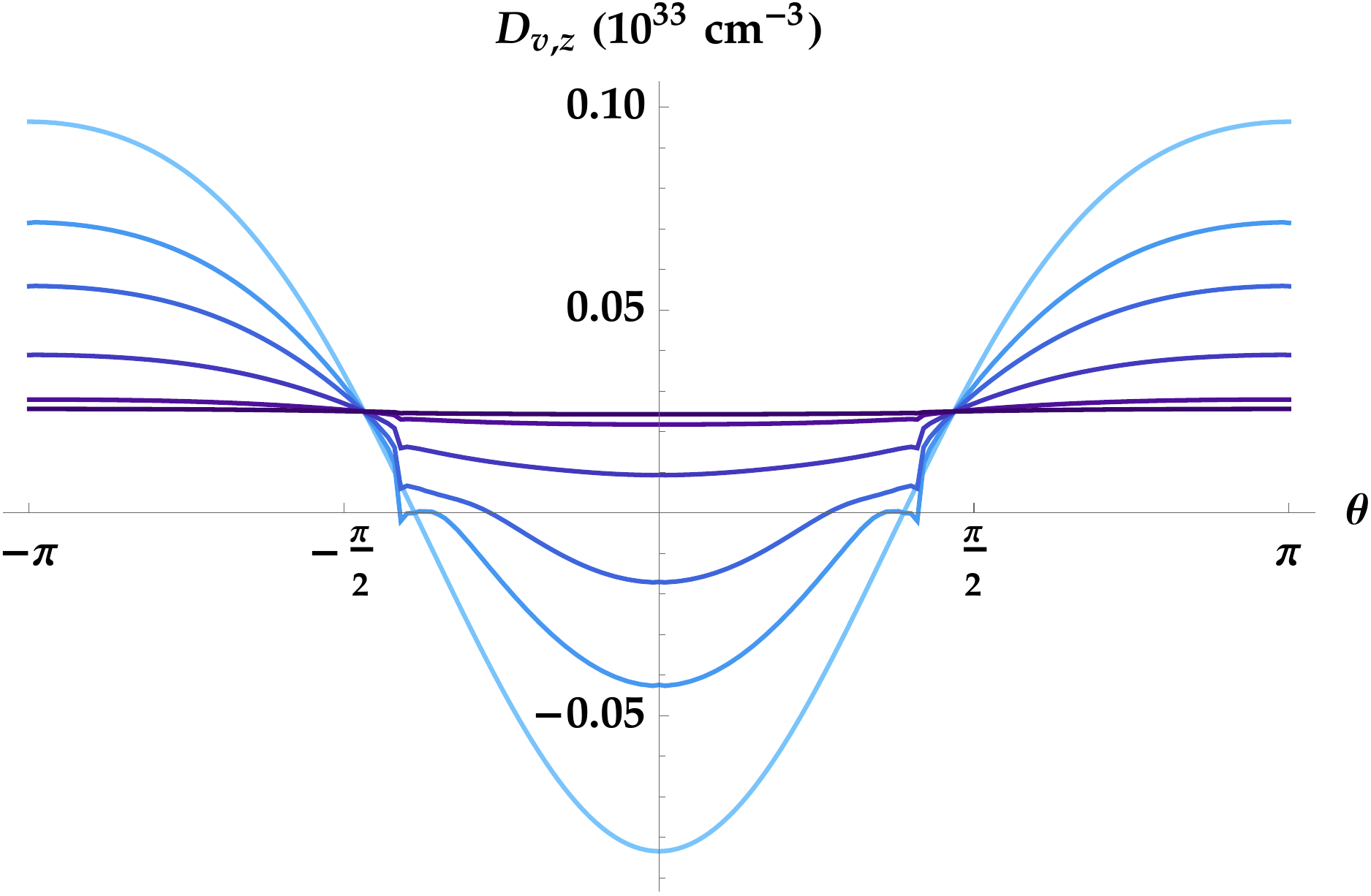}
}
\end{subfigure}
\begin{subfigure}{
\centering
\includegraphics[width=.310\textwidth]{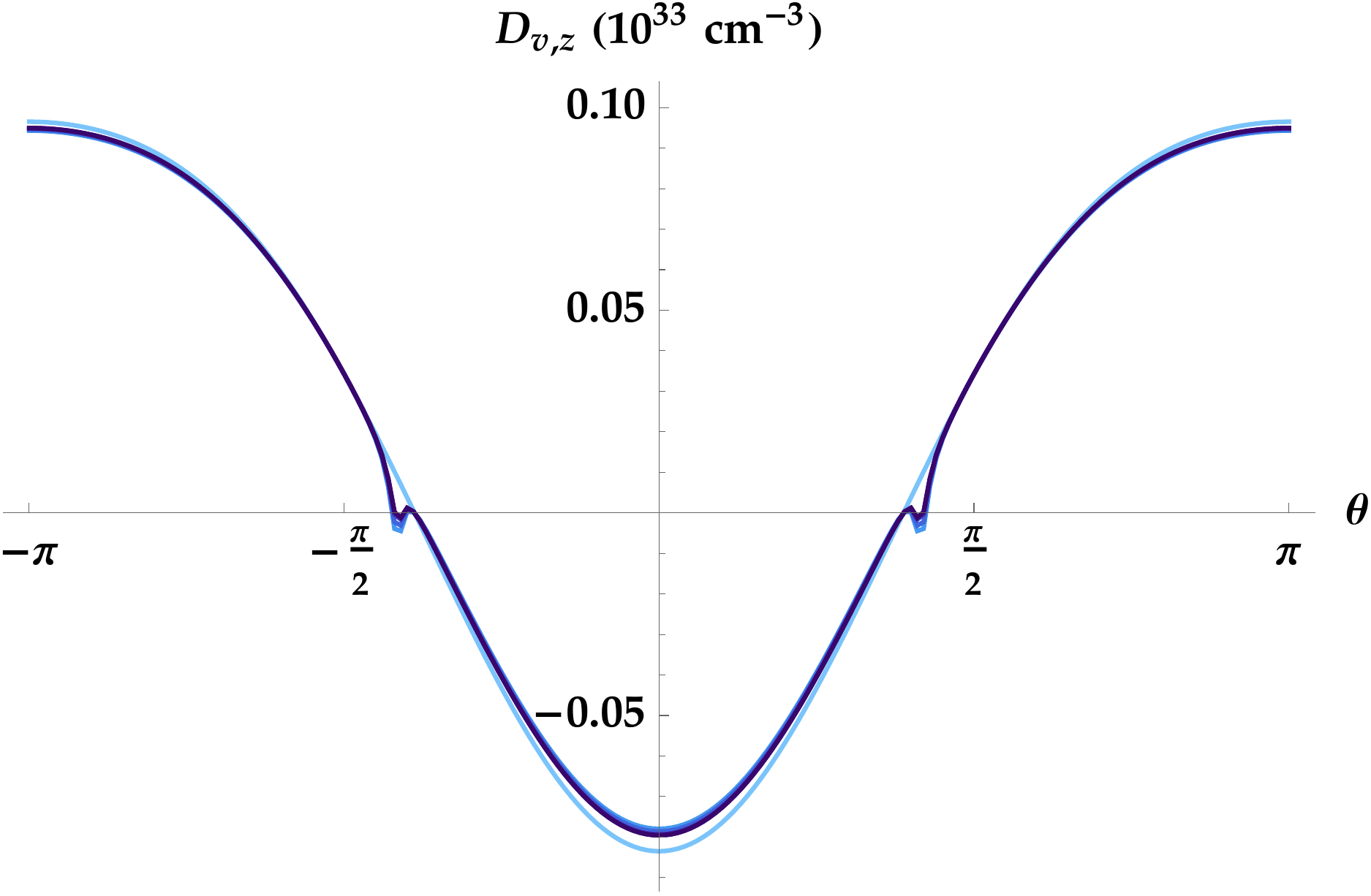}
}
\end{subfigure}

\begin{subfigure}{
\centering
\includegraphics[width=.310\textwidth]{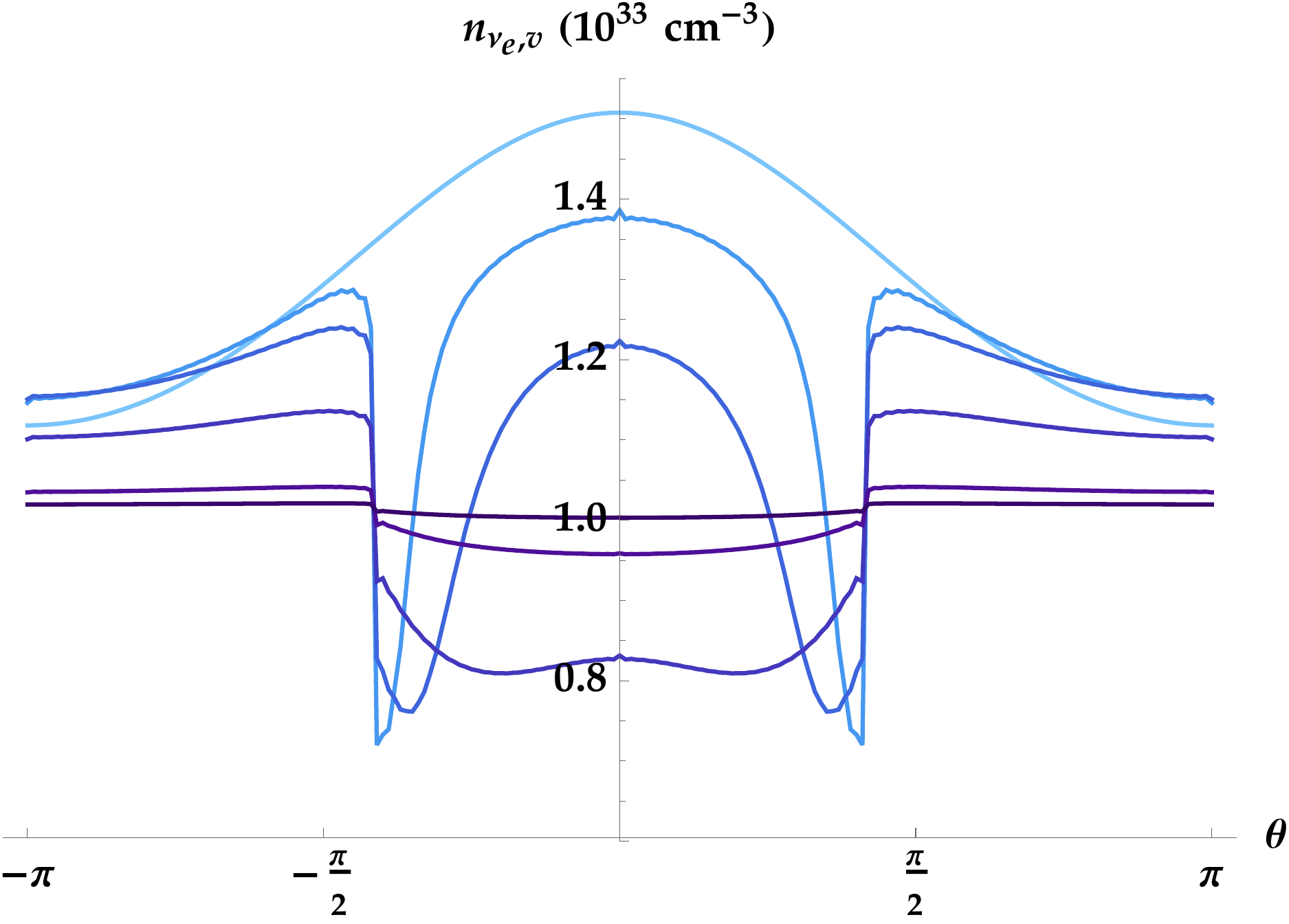}
}
\end{subfigure}
\begin{subfigure}{
\centering
\includegraphics[width=.310\textwidth]{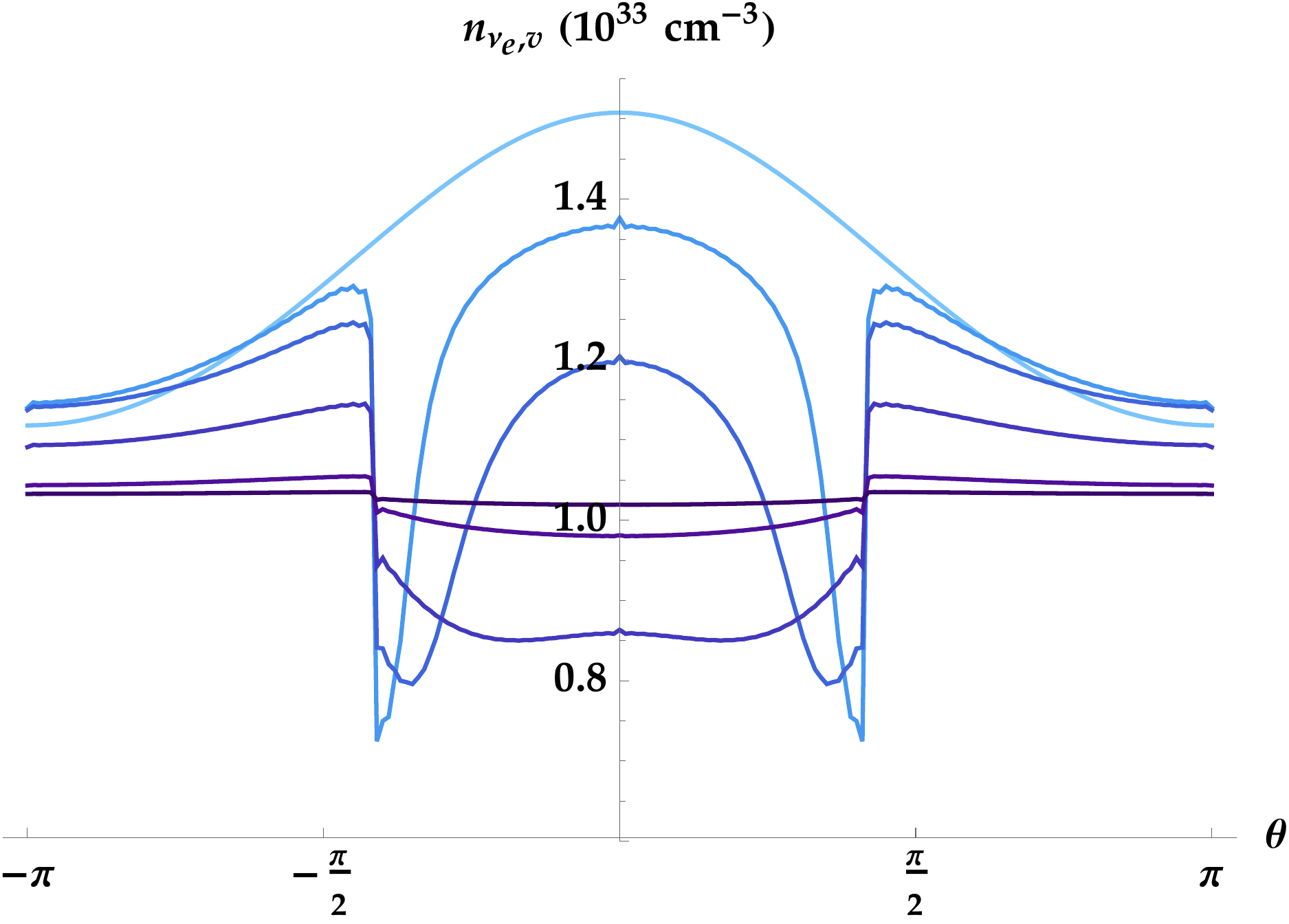}
}
\end{subfigure}
\begin{subfigure}{
\centering
\includegraphics[width=.310\textwidth]{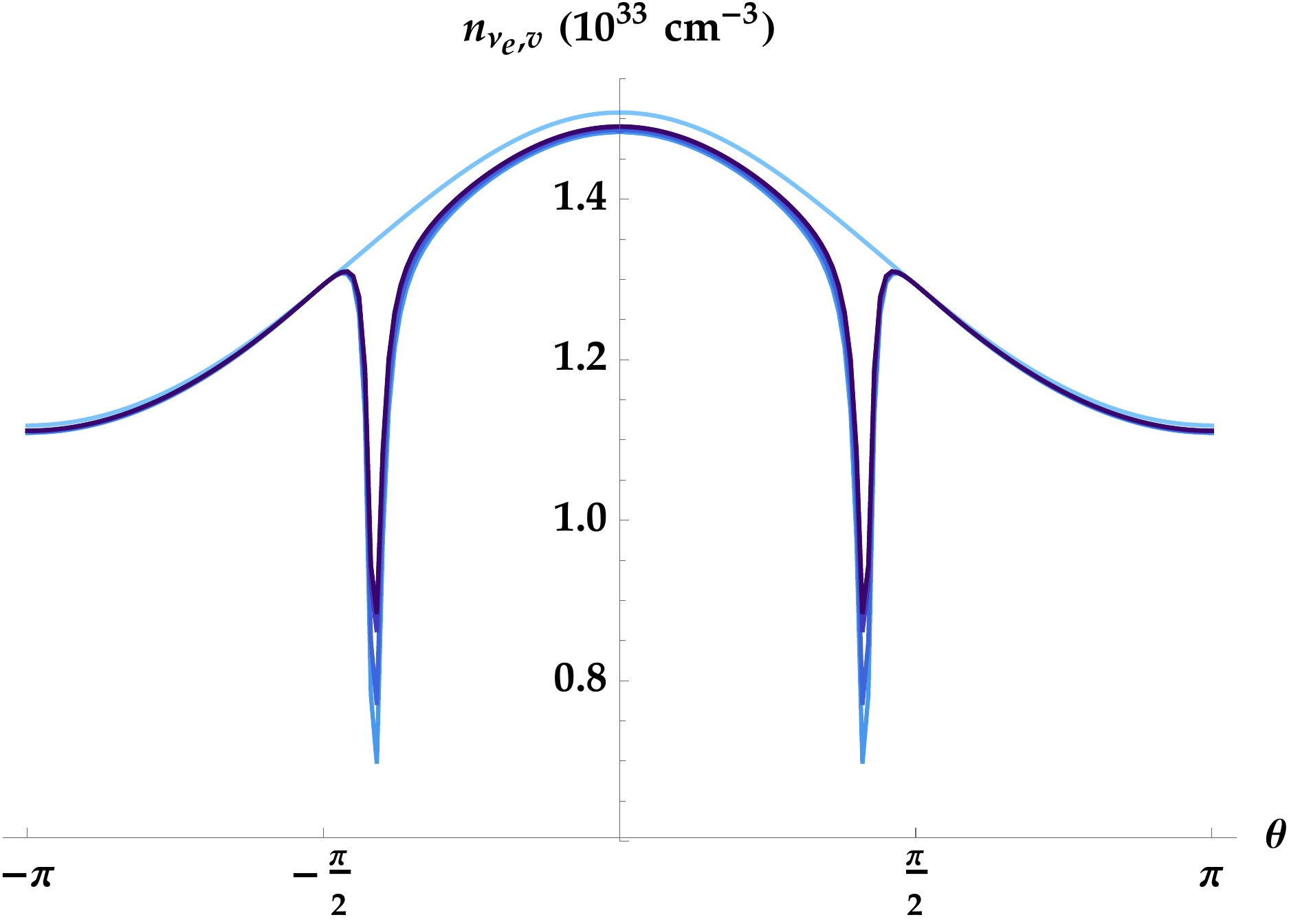}
}
\end{subfigure}
\caption{\textit{First row:} Evolution of the total number densities $n_{\nu_e}$ (top black curve), $n_{\bar{\nu}_e}$ (middle), $n_{\nu_x}$ (bottom), and neutrino coherence density $| \mathbf{P}_{0,T} | / 2$ (teal), where $T$ indicates the part of the vector transverse to the flavor axis. \textit{Second:} Evolution of $D_{1,z} / D_1$ (solid) and $-D_1 / D_1 (0)$ (dashed). \textit{Third:} Snapshots of $D_{v,z}$, where $v = \cos \theta$, at $t = 0$, $0.25$, $0.50$, $1.00$, $2.00$, and $3.00~\mu$s, color-coded from lightest to darkest. \textit{Fourth:} Snapshots of $n_{\nu_e, v}$ at the same times. The columns correspond to calculations with isotropizing NC scattering (\textit{left}), isotropizing CC scattering (\textit{middle}), and non-isotropizing CC scattering (\textit{right}). The collisionless case, not shown, is nearly identical to non-isotropizing CC, but without decoherence-induced reduction of the oscillation amplitudes. We interpret these results to mean that scattering-enhanced FFC is due to the isotropization of the angular distributions over time. Although scattering is isotropizing, neutrino angular distributions in a core-collapse supernova or neutron-star merger do not in fact isotropize in this manner, as discussed in the main text. We therefore conclude that the astrophysical relevance of scattering-enhanced FFC is at best ambiguous.}  
\label{figures}
\end{figure*}

Results of the calculations are presented in Fig.~\ref{figures}. The overall takeaways are as follows.

In agreement with other studies \cite{shalgar2021, kato2021, sasaki2021, hansen2022}, we find that NC scattering enhances the overall flavor conversion (top row, left panel). Without scattering, the initial fast oscillation---seen in the top row of Fig.~\ref{figures} as the rapid early-time variation with amplitude $\sim 0.1 \times 10^{33}~\textrm{cm}^{-3}$---simply repeats itself, without any reduction in amplitude, and the number densities periodically return to their initial values \cite{johns2020}. NC scattering causes these oscillations to damp away on a time scale of $\sim 0.2~\mu\textrm{s}$, and more significantly causes the number densities to deviate permanently from their starting points. For analysis of this behavior, we point to Refs.~\cite{sasaki2021} and \cite{hansen2022}.

Isotropizing CC scattering exhibits virtually identical enhancement (top row, middle panel). In fact, all important features of the dynamics are nearly identical between the two isotropizing forms of scattering, as evidenced by the left and middle columns. We see no phenomenology particular to neutral currents.

Non-isotropizing CC scattering shows markedly different results (top row, right panel). Again the oscillations damp away due to decoherence, but no additional flavor conversion takes place. FFC is enhanced only if the angular distributions isotropize as a function of time. As discussed previously, isotropization over time is precisely the feature of these calculations whose physical relevance is questionable. The enhancement effect itself is therefore of unclear relevance to natural environments.

The second row depicts the behavior of the monopole difference vector $\mathbf{D}_1$, which is the axially symmetric fast pendulum \cite{johns2020, padillagay2021}. It is constructed by defining polarization vectors $\mathbf{P}_v$ and $\mathbf{\bar{P}}_v$ with components satisfying
\begin{equation}
\rho_v = P_{v, 0} + \mathbf{P}_v, ~~~ \bar{\rho}_v = \bar{P}_{v, 0} + \mathbf{\bar{P}}_v,
\end{equation}
where the traces $2P_{v,0}$ and $2\bar{P}_{v,0}$ give the number densities summed over the two flavors. Let subscript $l$ denote the $l$th Legendre moment, \textit{e.g.},
\begin{equation}
\rho_{l} = \int_{-1}^{+1} dv L_l(v) \rho_v,
\end{equation}
where $L_l (v)$ is the $l$th Legendre polynomial, and similarly for other quantities. Then the fast pendulum is the $l = 1$ moment of the difference vector $\mathbf{D}_v = \mathbf{P}_v - \mathbf{\bar{P}}_v$, \textit{i.e.},
\begin{equation}
\mathbf{D}_1 = \int_{-1}^{+1} dv ~v ( \mathbf{P}_v - \mathbf{\bar{P}}_v ),
\end{equation}
and $D_{1,z}$ is its component along the flavor axis. From the figure we see that all cases begin with small-amplitude oscillations corresponding to the swinging of the fast pendulum. Decoherence (right panel) pushes $\mathbf{D}_1$ back along the $\mathbf{z}$ axis, pinning it there. Isotropization (left and middle panels) causes the vector to decay while maintaining a nonzero angle between the $\mathbf{z}$ axis and itself.

For reference, we also explicitly define the $l=0$ difference vector
\begin{equation}
\mathbf{D}_0 = \int_{-1}^{+1} dv ( \mathbf{P}_v - \mathbf{\bar{P}}_v ), \label{eq:D0}
\end{equation}
which sets the overall asymmetry between $\nu_e$ and $\bar{\nu}_e$. This vector is used in Table~\ref{tablevparams}.

In the third row we see $D_{v,z}$ at several moments in time. Decoherence on its own has little effect, but isotropization causes the ELN crossings to move inward toward $\theta = 0$. At late times $D_{v,z}$ has flattened and the crossings have disappeared.

The bottom row plots the number densities $n_{\nu_e, v}$ at the same snapshots. With the chosen parameters, collisionless FFC exhibits narrow dips, which are seen here in the bottom row's right panel. These features are explained in Ref.~\cite{johns2020}: the dips encompass trajectories whose flavor states are in a kind of resonance with $\mathbf{D}_1$. Under the influence of decoherence, but without isotropization, the FFC dips are frozen in place with a small reduction in depth. Scattering in general causes flavor conversion to become locked in. If scattering isotropizes as well as decoheres (bottom row, left and middle panels), the initially narrow dips spread inward as the ELN crossings drift toward each other. A trough forms that is centered at $\theta = 0$, a cumulative effect of the ELN inciting fast conversion that is then frozen in by decoherence. Scattering redistributes flavor over all angles, causing $n_{\nu_e,v}$ to flatten out, but the enhancement of total flavor conversion (top row, left and middle panels) comes from the cumulative effect just described.

\section{Summary \label{sec:discussion}}

This work was inspired by the apparent discrepancy in the findings of Refs.~\cite{shalgar2021, martin2021, sigl2022, kato2021, sasaki2021, hansen2022} concerning whether scattering enhances or suppresses FFC. We have made three main points.

Firstly, calculations that combine spatial homogeneity, anisotropic angular distributions, and collisional flavor evolution are subject to interpretive issues because the anisotropic distributions are not generated organically. Starting a calculation with anisotropic distributions, which then isotropize due to collisions, is not self-consistent. By our definition, self-consistency means that the initial state of the flavor field is constant under the classical equations of motion. This is not always a pertinent definition, of course, but it is useful here because it prevents the conflation of collective flavor phenomenology and transient behavior unrelated to flavor mixing. In a model that is self-consistent by this criterion, any new effect that appears when the Hamiltonian is turned on is a genuinely new flavor-mixing phenomenon. Collisionless instabilities are clearly self-consistent in this sense. Less obviously, so are collisional instabilities, which manifest in isotropic models \cite{johns2021b}.

Secondly, in our anisotropic model, we find no significant differences between the scenarios with isotropizing NC and CC scattering. The flavor-blind character of neutral currents is not required for scattering-enhanced FFC to occur.

Lastly, scattering-enhanced FFC is due to isotropization of the angular distributions, which shifts the location of the ELN crossing and eventually causes the crossing to vanish. This finding implies that the enhancement may not be an astrophysically relevant effect, since isotropization of this kind is not a realistic feature of compact-object environments.

Given some location in a supernova or merger, ELN crossings (regarded as a function of propagation angle) do not in fact drift and disappear in the manner observed in our isotropizing calculations. While crossings do change over time, in reality they do so because the physical conditions of the environment are changing. That phenomenon is wholly separate from what is observed here and in Refs.~ \cite{shalgar2021, kato2021, sasaki2021, hansen2022}.

On the other hand, there may be enhancement effects \textit{analogous} to the one we have focused on. On the time scale of collective oscillations, angular distributions are nearly constant, but they of course change as a function of position. Flavor conversion in one region can convect into another region where, if the problem is considered locally, one expects different flavor transformation. In this sense there could perhaps be a kind of cumulative effect like the one described in Sec.~\ref{sec:results}, with position substituting for time. But, notwithstanding the similarity, this effect is again distinct from scattering-enhanced FFC.

The self-consistency issues surrounding isotropization can be avoided by performing calculations where the angular distributions \textit{are} formed organically by advection and collisions, using a model that is inhomogeneous in neutrinos and matter. Sigl \cite{sigl2022} has recently run inhomogeneous calculations and found no enhancement of FFC. In a section discussing the numerical results, he speculated on the reason for the discrepancy between his findings and those of Refs.~\cite{shalgar2021} and \cite{sasaki2021}, commenting that it may have to do with the latter studies' assumptions of homogeneity. Our own findings support and expand on that remark.

\begin{acknowledgments}
The authors thank Baha Balantekin, Basudeb Dasgupta, Amol Dighe, George Fuller, Evan Grohs, Chinami Kato, Jim Kneller, Gail McLaughlin, Amol Patwardhan, Sherwood Richers, and Zewei Xiong for valuable discussions about this work and related topics. Support for this work was provided by NASA through the NASA Hubble Fellowship grant number HST-HF2-51461.001-A awarded by the Space Telescope Science Institute, which is operated by the Association of Universities for Research in Astronomy, Incorporated, under NASA contract NAS5-26555.
\end{acknowledgments}

\bibliography{all_papers}

\end{document}